\documentclass{emulateapj}
\usepackage{amsmath}
\usepackage{amssymb}
\usepackage{graphicx}
\usepackage{appendix}
\usepackage{color}
\usepackage{multirow}
\usepackage{rotating}
\newcommand{\lSect}[1]{{\label{sec:#1}}}
\newcommand{\lFig}[1]{{\label{fig:#1}}}

\newcommand{\lTab}[1]{{\label{tab:#1}}}
\newcommand{\FIGFF}[2]{{\ref{fig:#2}{#1}}}

\newcommand{\FIG}[2]{{Fig.~\FIGFF{#1}{#2}}}
\newcommand{\Fig}[1]{{\FIG{}{#1}}}

\newcommand{\Msun}{\ensuremath{\mathrm{M}_\odot}}

\newcommand{\Zsun}{\ensuremath{\mathrm{Z}_{\odot}}}
\newcommand{\Tab}[1]{{Table \ref{tab:#1}}}

\newcommand{\h}[1]{{\hspace{#1}}}
\newcommand{\vl}{\vline}

\def\gtaprx {\lower .1ex\hbox{\rlap{\raise .6ex\hbox{\hskip .3ex
	{\ifmmode{\scriptscriptstyle >}\else
		{$\scriptscriptstyle >$}\fi}}}
	\kern -.4ex{\ifmmode{\scriptscriptstyle \sim}\else
		{$\scriptscriptstyle\sim$}\fi}}}
\def\ltaprx {\lower .1ex\hbox{\rlap{\raise .6ex\hbox{\hskip .3ex
	{\ifmmode{\scriptscriptstyle <}\else
		{$\scriptscriptstyle <$}\fi}}}
	\kern -.4ex{\ifmmode{\scriptscriptstyle \sim}\else
		{$\scriptscriptstyle\sim$}\fi}}}

\begin{document}

\title{The Progenitor of GW 150914}

\author{S. E. Woosley\altaffilmark{1}}

\altaffiltext{1}{Department of Astronomy and Astrophysics, University
  of California, Santa Cruz, CA 95064; woosley@ucolick.org}

\begin{abstract} 
The spectacular detection of gravitational waves (GWs) from GW 150914
and its reported association with a gamma-ray burst (GRB) offer new
insights into the evolution of massive stars. Here it is shown that no
single star of any mass and credible metallicity is likely to produce
the observed GW signal. Stars with helium cores in the mass range 35
to 133 \Msun \ encounter the pair instability and either explode or
pulse until the core mass is less than 40 \Msun, smaller than the
combined mass of the observed black holes. The rotation of more
massive helium cores is either braked by interaction with a slowly
rotating hydrogen envelope, if one is present, or by mass loss, if one
is not.  The very short interval between the GW signal and the
observed onset of the putative GRB in GW 150914 is also too short to
have come from a single star. A more probable model for making the
gravitational radiation is the delayed merger of two black holes made
by 70 and 90 \Msun \ stars in a binary system. The more massive
component was a pulsational-pair instability supernova before making
the first black hole.
\end{abstract}


\section{INTRODUCTION}
\lSect{intro}

The detection by \citet{Abb16} of gravitational radiation from a pair
of merging massive black holes, $36_{-4}^{+5}$ and $29_{-4}^{+4}$
\Msun, has ushered in a new era of multi-messenger astronomy.
Motivated largely by the reported coincident gamma-ray burst
\citep{Con16}, \citet{Loe16} suggested that a single star model might
have produced GW 150914.  He assumes the precursor to the collapse
contains a rapidly rotating helium core of at least 65 \Msun \ (the
sum of the masses of the observed black hole pair) that has formed as
the result of a prior merger of stars in a binary system. The core
rotates so rapidly that it fissions into two black holes during its
collapse. The two back holes recombine emitting the observed GW.

Well before the GW detection, \citet{Fry01} calculated an alternative,
single star model for a burst of gravitational radiation accompanied
by a GRB, but invoked a much more massive star, 300 \Msun, with a 180
\Msun \ helium core. Their 2D calculation was unable to directly
demonstrate the fission of the collapsing core, but rotational energy
and angular momentum considerations suggested that as a possible
outcome.

Here it is shown that both of these scenarios are unlikely to produce
the observed GW event. Helium cores anywhere near 65 \Msun, and up to
133 \Msun, encounter the pair instability, and either pulse violently
and lose mass \citep{Woo07,Woo15}, or explode \citep{Heg02}. They do
not make black holes having a total mass greater than about 40
\Msun. For helium cores above 133 \Msun \ black hole production is
indeed likely, but when reasonable estimates of magnetic torques and
mass loss are included \citep{Heg05}, the collapsing core contains
too little angular momentum to bifurcate.

Any stellar channel for producing GW 150914 must therefore involve
binary evolution. Here it is shown that the observationally inferred
masses for the merging black holes, $36_{-4}^{+5}$ and $29_{-4}^{+4}$
\Msun, could result from the evolution of stars of non-rotating stars
of 70 and 90 \Msun \ with metallicity less than about 10\% solar. If
rotation is included, the inferred masses are closer to 60 and 70
\Msun. Interestingly, the heavier star of each pair not only produces
a black hole near 36 \Msun, but also makes a pulsational-pair
instability supernova (PPISN) that ejects any low density envelope
shortly before collapse, but does not unbind the system.

\begin{deluxetable*}{rccccll} 
\tablecaption{LOW METALLICITY MODELS} 
\tablehead{  & \vspace{-0.1cm}
                     & 
                    $\rm M_{ZAMS}\ \vl\ M_{preSN}\ \vl\ M_{remnant}$ & 
                    $\rm M_\alpha\ \vl\ M_{CO}\ \vl\ M_{Si}\ \vl\ M_{Fe}$ & 
                    Pulse Activity \\ \vspace{-0.1cm}
                    Models &
                    Mass Loss &
                    &&\\
                    &  
                    & 
                    [\Msun] & 
                    [\Msun] & 
                    $\hspace{1.4em}\rm t_p\ [10^6\ s]\ \vl\ E_p\ [10^{50}\ ergs]$}
\startdata
\parbox[t]{4mm}{\multirow{5}{*}{\rotatebox[origin=c]{90}{no rotation}}}
T70A                     & 1 & 70   $\vl$ 55.5 $\vl$ 55.5                     & 30.1 $\vl$ 26.4 $\vl$ 7.87 $\vl$ 2.58           & 0.059 $\vl$ -    $\h{2.3em}$\\
T70B                     & 0 & 70   $\vl$ 70.0 $\vl$ 70 $\h{0.45em}$& 31.6 $\vl$ 28.0 $\vl$ 8.41 $\vl$ 2.57           & 0.055 $\vl$ -    $\h{2.3em}$\\
T90A                     & 1 & 90   $\vl$ 66.5 $\vl$ 35.9                     & 39.7 $\vl$ 35.4 $\vl$ 9.54 $\vl$ 2.57           &     1.1 $\vl$ 5.6$\h{0.69em}$\\
T90B                     & 0 & 90   $\vl$ 90.0 $\vl$ 37.1                     & 40.9 $\vl$ 36.8 $\vl$ 8.35 $\vl$ 2.86           &     1.9 $\vl$ 4.9$\h{0.69em}$\\
T150$\h{0.2em}$& 1 & 150 $\vl$ 96.0 $\vl$ 0   $\h{1.45em}$& 65.1 $\vl$ 55.2 $\vl$ - PSN - $\h{0.62em}$&         - $\vl$ 77$\h{0.02em}$                      \\
\vspace{-0.2cm}\\
\hline\vspace{-0.2cm}
&&&&&\\
\parbox[t]{4mm}{\multirow{9}{*}{\rotatebox[origin=c]{90}{rotating}}}
R60    $\h{0.3em}$ &1/2 &60   $\vl$                 52.4 $\vl$ 52.3                     &30.9 $\vl$ 26.6 $\vl$ 8.18  $\vl$ 2.64                          & 0.047 $\vl$ -    $\h{2.3em}$\\
R70    $\h{0.3em}$ &1    &70   $\vl$                 56.9 $\vl$ 34.5                     &37.4 $\vl$ 32.9 $\vl$ 10.3  $\vl$ 2.72                          & 0.32   $\vl$ 4.5$\h{1.2em}$\\
R110  $\h{-0.2em}$&1/2 &110 $\vl\h{0.62em}$102 $\vl$ 0$\h{1.78em}$  &66.7 $\vl$ 60.3 $\vl$ - PSN - $\h{0.62em}$                & -         $\vl$ 62$\h{0.03em}$\\
R150A$\h{-0.6em}$&0    &150 $\vl\h{0.62em}$150  $\vl$ 150$\h{0.78em}$ &150  $\vl\h{2.6em}$ - BH - $\h{0.76em}$                    & -         $\vl$ -$\h{0.7em}$      \\
R150B$\h{-0.6em}$&1    &150 $\vl$                 28.4 $\vl$ 28.4$\h{0.5em}$&28.4 $\vl$ 24.3 $\vl$ 6.55  $\vl$ 2.58                          & -         $\vl$ -$\h{0.7em}$      \\
R150C$\h{-0.6em}$&1    &150 $\vl\h{0.62em}$122  $\vl$ 0$\h{1.78em}$     &122  $\vl\h{0.28em}$ 118 $\vl$ - PSN - $\h{0.35em}$& -         $\vl$ 768$\h{-0.46em}$\\
R300A$\h{-0.6em}$&0    &300 $\vl\h{0.62em}$300  $\vl$ 300$\h{0.78em}$ &161  $\vl\h{2.6em}$ - BH - $\h{0.77em}$                    & -         $\vl$ -$\h{0.7em}$      \\
R300B$\h{-0.6em}$&0    &300 $\vl\h{0.62em}$300  $\vl$ 300$\h{0.78em}$ &180  $\vl\h{2.6em}$ - BH - $\h{0.77em}$                    & -         $\vl$ -$\h{0.7em}$      \\
R300C$\h{-0.6em}$&1    &300 $\vl\h{0.62em}$143  $\vl$ 143$\h{0.78em}$ &143  $\vl\h{2.6em}$ - BH - $\h{0.77em}$                    & -         $\vl$ -$\h{0.7em}$     \\
\vspace{-0.2cm}
\enddata
\lTab{tmodels}
\end{deluxetable*}

\section{THE FAILURE OF SINGLE STAR MODELS FOR GW 150914}
\lSect{single}

Since the hydrogen envelope will not participate in any prompt
collapse, a minimum helium core mass equal to the sum of the observed
black hole masses is required in any single star model, i.e.,
$M_{\alpha} > 65$). Since it is unlikely that the entire helium core
collapsed during the less than 1 second duration of the GW signal,
this is a lower bound, possibly an extreme one \citep{Woo86}. Such a
large helium core could be the consequence of a single star of mass
over 150 \Msun, the merger of two lighter stars, or the chemically
homogeneous evolution evolution of a star as small as 65 \Msun.

To guide the discussion, a set of models (Table 1) was calculated
using the \texttt{KEPLER} code \citep[e.g.]{Wea78,Woo02}. The models
had in common a low metallicity, 10\% that of the sun, chiefly
employed to keep the mass loss rate low. Most of the models that
included rotation also included magnetic torques \citep{Heg05} that
acted to brake the rotation of the helium core at late times and
enforce rigid rotation. Modern, but uncertain mass loss rates were
included \citep{Nie90}. For hydrogen burning stars with surface
hydrogen mass fractions more than 0.4, a metallicity scaling of
Z$^{0.5}$ was employed. For chemically homogeneous models with lower
surface hydrogen mass fractions, the Wolf-Rayet mass loss rate of
\citet{Bra97} was employed as modified for clumping by \citet{Woo06}
with a metallicity scaling of Z$^{0.86}$ \citep{Vin05}. All stars
considered would have lost their hydrogen envelopes and most of their
helium cores had they been of solar metallicity.

Models starting with ``T'' in Table 1 were not rotating; those
starting with ``R'' included rotation. One non-rotating model, T150,
had a final helium core mass of 65 \Msun \ and illustrates that helium
cores of this mass explode as ordinary PISN. They do not make black
holes. Including rotation is not likely to alter this outcome
\cite{Cha13,Che15}, though it does reduce the main sequence mass
required to produce the helium core (Model R110).

Two other sets of models were calculated that included
rotation. Models R300 were based on the evolution of single stars of
10\% solar metallicity with initial masses of 300 \Msun. Each had a
total initial angular momentum of $1.5 \times 10^{54}$ erg s and an
equatorial rotational speed of 180 km s$^{-1}$. Model R300A did not
include mass loss or magnetic torques, and is thus very similar to the
case studied by \citet{Fry01}. Model R300B included magnetic torques
but not mass loss; R300C included both mass loss and magnetic torques.
Models R150 are discussed later.

\begin{figure}[h]
\centering \includegraphics[width=\columnwidth]{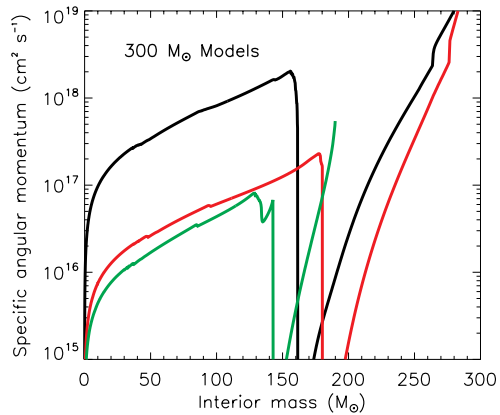}
\caption{Final angular momentum distribution at collapse of a 300
  \Msun \ model evolved with no mass loss or magnetic torques (black
  line); magnetic torques, but no mass loss (red line); and mass loss
  plus magnetic torques (green line). The edge of the helium core is
  apparent where the angular momentum sharply drops and then rises
  again. Both models that include magnetic torques have too little
  angular momentum for the core to bifurcate or form a disk during its
  collapse to a black hole. The Kerr parameter is of order unity for
  the black curve, but everywhere less than 0.1 for the green and red
  curves. The helium cores of all three models are sufficiently
  massive to collapse directly to black holes. \lFig{r300}}
\end{figure}

\Fig{r300} shows the distribution of specific angular momentum, j, in
the three 300 \Msun \ models at the point of pair instability. The
central temperature in each exceeded $3 \times 10^9$ K and no further
redistribution of j will occur.  The helium core here is about 10\% less
than that of \citet{Fry01} because convective dredge up was less
efficient in the new calculation, but the angular momentum
distribution is quite similar (compare the black line in \Fig{r300}
with their Fig. 2). The total angular momentum of the helium core
is $2.36 \times 10^{53}$ erg s. As \citet{Fry01} noted, this is
sufficient angular momentum to form a disk, and perhaps even to
bifurcate and emit copious gravitational radiation in its final stage
of collapse.

Unfortunately, as \Fig{r300} also shows, the inclusion of magnetic
torques and mass loss greatly alters the outcome. The helium core
still exceeds the critical mass, 133 \Msun, for producing a black
hole, but its angular momentum is greatly reduced by interacting with
the slowly rotating hydrogen envelope. The angular momentum in the
helium core of R300B at collapse is reduced by a factor of 10 to $2.48
\times 10^{52}$ erg s, and inadequate even to make a disk anywhere
except perhaps in the outermost core. Model R300C, with torques and
mass loss, is worse, with a total angular momentum in the helium core
of only $8.74 \times 10^{51}$ erg s and insufficient angular momentum to
make a disk anywhere in the core.  Models R300B and 300C, which are
more realistic than R300A, will not make the observed GW signal.

One might object that the mass loss rates and especially the
prescription for magnetic torques are uncertain. This is true, but the
magnetic torques used here give approximately the right rotation rates
for newly born pulsars \citep{Heg05} and, if anything, might need to
be larger. Neglecting them leads to most massive stars dying with
sufficient angular momentum to make a millisecond magnetar. The fact
that gamma-ray bursts are so rare, is thus an argument that magnetic
torques of a magnitude not too different from that assumed here must
play a role in the evolution of massive stars.

One way to give the core additional angular momentum might be to avoid
the production of any red supergiant phase and invoke chemically
homogeneous evolution (CHE) on the main sequence
\citep{Woo06,Yoo12,Cha12,Koh15}. This is the tack taken by those
attempting to explain the rapid rotation of gamma-ray burst
progenitors, and would presumably correspond to a rare event limited
to stare of unusually rapid rotation. To test this possibility, three
more models were calculated with a lower mass, but greater specific
angular momentum to insure CHE. Models R150, with mass 150 \Msun, had
an initial total angular momentum of $1 \times 10^{54}$ erg s and
rotated at 325 km s$^{-1}$ on the main sequence. Each experienced CHE
and included magnetic torques. Model R150A had no mass loss though,
while Models R150B and R150C had a mass loss rate equal to the full
value or 10\% of the full value expected for a star with metallicity
10\% that of the sun.

\begin{figure}[h]
\centering \includegraphics[width=\columnwidth]{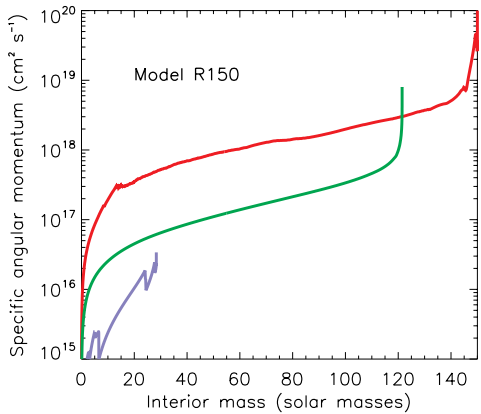}
\caption{Final angular momentum distribution in a 150 \Msun \ model
  evolved including sufficient angular momentum that it experiences
  complete mixing on the main sequence. Model R150A (top red line) had
  no mass loss. Models R150B (bottom blue line) and R150C (middle
  green line) included 100\% and 10\% respectively of the mass loss
  expected for a star with 10\% solar metallicity. \lFig{r150}}
\end{figure}

The results for Model R150A are encouraging.  By the time the central
density reached 10$^{11}$ g cm$^{-3}$ when neutrino trapping might
occur, the ratio of rotational energy to gravitational potential
energy, T/W, exceeded 20\% in the inner 10 \Msun, and was greater than
12\% everywhere in the star.  The Kerr parameter was substantially
greater than unity throughout the entire core, guaranteeing that
angular momentum will be lost, either through a disk or fission and
gravitational radiation, before even a maximally rotating black hole
forms. The subsequent evolution would probably be similar to that of
the 180 \Msun \ core of \citet[][Compare with their Fig. 3]{Fry01} in
2D and, possibly, to the calculations of \citet{Rei13} in 3D.

Unfortunately, this is not a credible model since mass loss must be
included. If the mass loss appropriate to a 10\% solar metallicity
star is used, the star nearly evaporates (\Fig{r150}). Not only is the
final mass too small to make the pair of black holes, the amount
of residual angular momentum is trivial.

Mass loss rates are uncertain and perhaps the metallicity was even lower
than 0.1 \Zsun, so model R130C used ten times less mass loss. This
would make the model comparable to those studied by \citet{Koh15} for
massive stars in the galaxy I Zwicky 18 with metallicity 10$^{-1.7}$
that of the sun. Indeed, at hydrogen depletion, the remaining mass for
Model R150C here, 139 \Msun, is close to what they calculated for a star of
similar rotation rate, 136 \Msun. Much more mass loss occurs during
helium burning though, and the star ends up with a final mass of 122
\Msun. This is too small to produce a black hole, but that deficiency
might be alleviated by taking a more massive main sequence star. More
problematic is the low angular momentum in the star at death, $5.0
\times 10^{52}$ erg s. While adequate to produce a disk in the outer
core if a black hole formed, the angular momentum in the inner core is
too little to cause fission.

Thus single star models that include magnetic torques and mass loss
fail to produce a system that could explain GW 150914. If an envelope
is present the core is braked too much by the interaction. If the
envelope is absent, mass loss from the core has the same consequence.

\begin{figure}
        \centering
        \includegraphics[width=0.8\columnwidth]{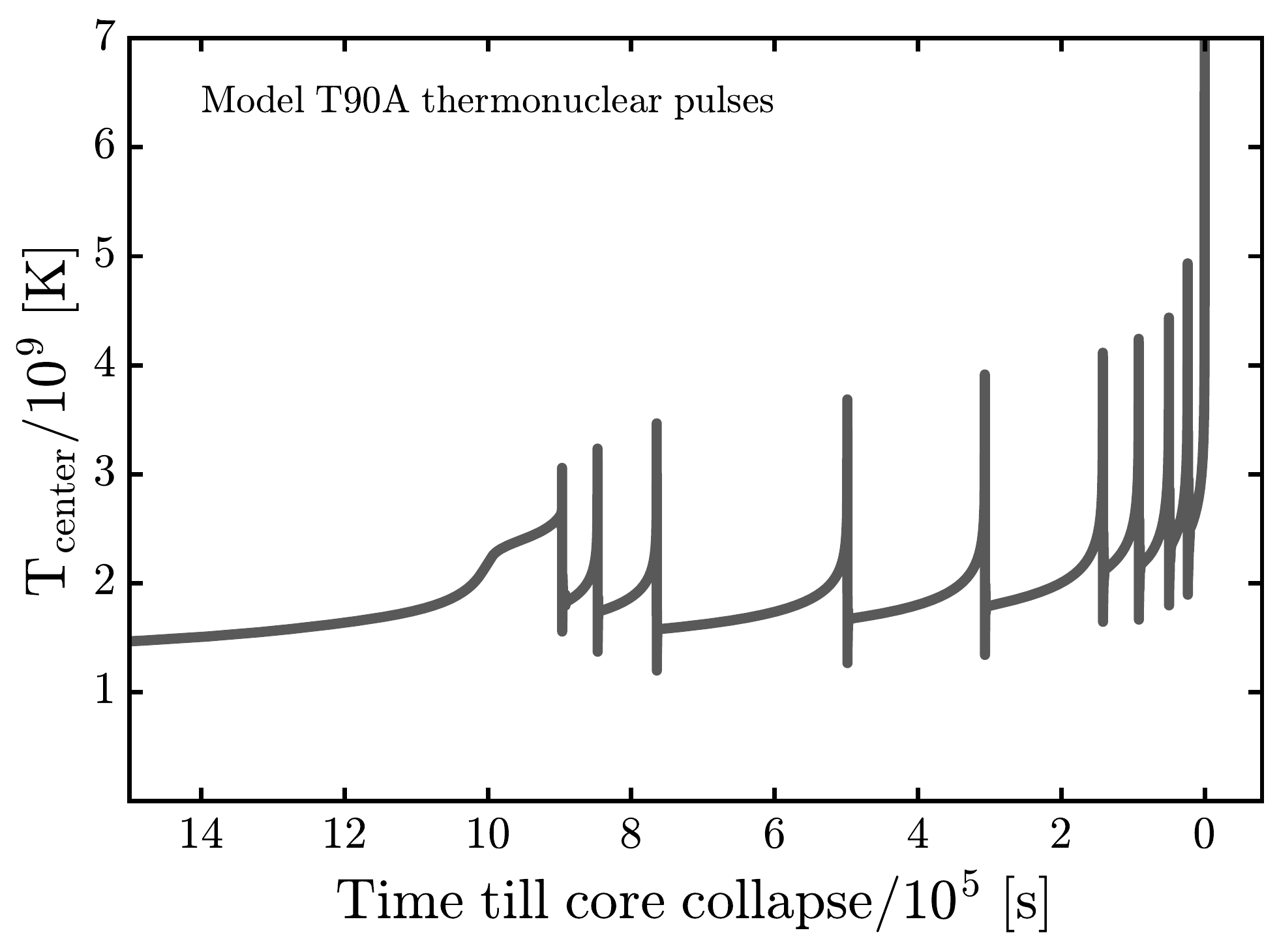}
        \includegraphics[width=0.8\columnwidth]{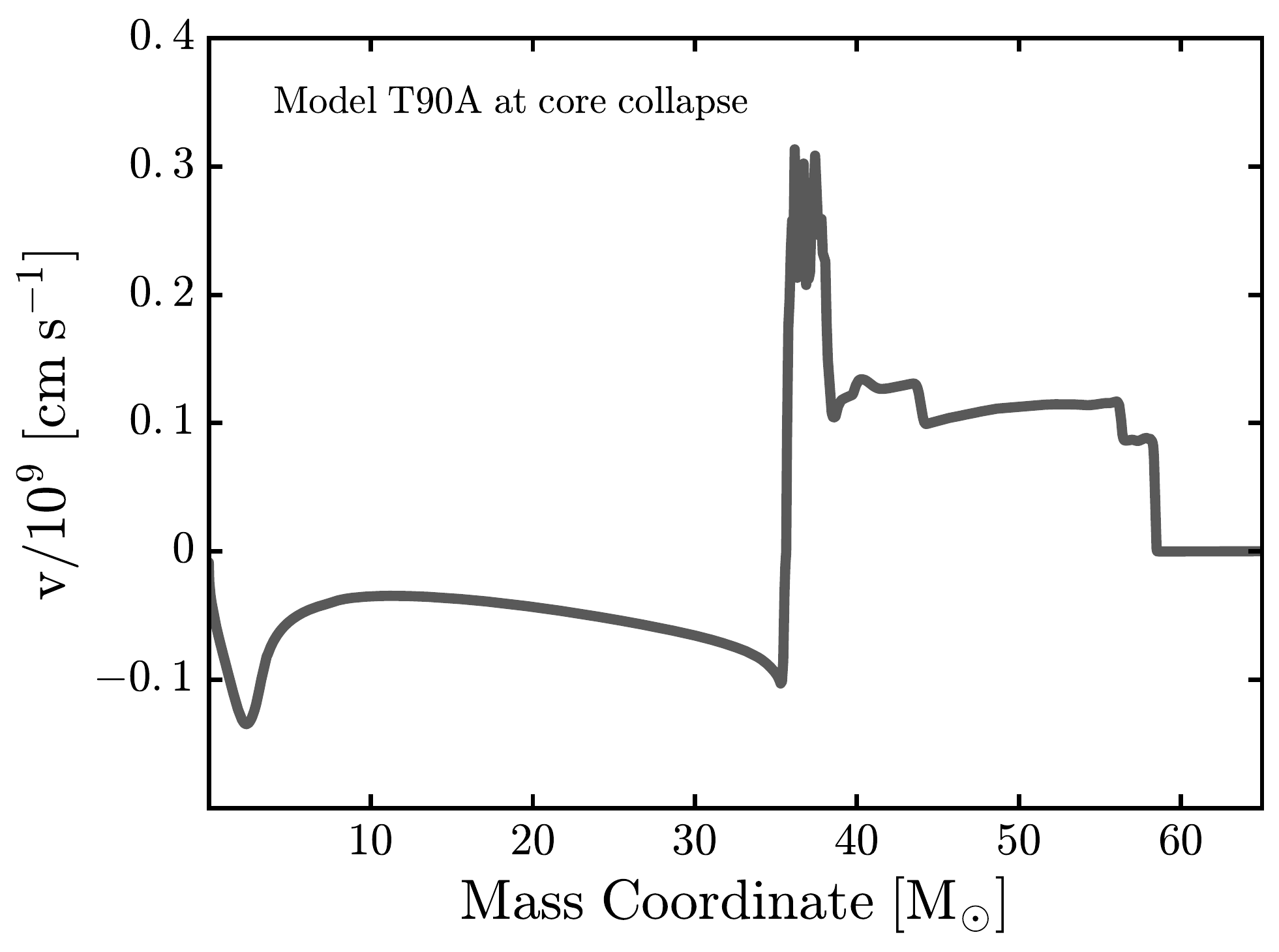}
        \caption{Pulse history and collapse velocity for Model T90A.
  \lFig{t90coll}}
\end{figure}

\section{A BINARY MODEL FOR GW 150914}
\lSect{binary}

Alternatively, as suggested many times before the event was observed,
\citep[e.g.]{Bel10,Bel16a,Mar16,Man16}, the two black holes could be
made separately in a binary system and brought together by a
combination of common envelope evolution and gravitational radiation.

To provide some detail to these suggestions, which did not
include the evolution of the stars to their collapse, several other
models were calculated (Table 1). The main sequence masses here were
chosen to produce helium cores in the resulting presupernova star near
29 (Models T70 and R60) and 36 \Msun \ (Models T90 and R70). Models
T70 and T90 did not include rotation. Models R60, R70, and R110 had an
equatorial rotation speed on the main sequence of 160 - 180 km
s$^{-1}$ corresponding to a ratio of centrifugal force to gravity of
approximately 15\%.  These models did not experience chemically
homogeneous evolution, and their helium core masses were substantially
less than their final masses. All models ended their (single star)
lives as red supergiants.  Since mass loss is uncertain, some models
were run both with and without mass loss (A and B). The results show
that the dependency of final helium core mass on the mass loss rate is
weak, so long as the entire envelope is not lost. For those models
that experienced the pulsational pair instability, the number of
pulses and total energy in pulses are given. No pulse energies are given
for Models T70 and R60 because the they were very weak and the
envelope was not ejected.

\subsection{The Components}

Model T90A, which is exemplary of stars that make a 36 \Msun \ black
hole, encountered the pulsational-pair instability at carbon
depletion. This resulted in nine strong thermonuclear pulses
occurring during the last 10$^6$ s of the star's life
(\Fig{t90coll}). Following each flash, the core expanded, emitted a shock
wave into the hydrogen envelope, then relaxed, and encountered the pair
instability again.  These flashes derived their energy from explosive
oxygen and silicon burning in the inner 5 \Msun \ of the
core. Eventually the pulses ejected the envelope leaving most of the
helium core intact and stable against further pulsation. The residual
core evolved to produce an iron core of 2.57 \Msun \ (Model T90A) or
2.86 \Msun \ (Model T90B).  While the subsequent evolution of the iron
core was not followed here, given its large size and the shallow
density gradient outside, it is unlikely to explode by neutrino
transport \citep[e.g.]{Ert16}. Lacking rotation, the protoneutron star
accretes for perhaps a second and becomes a black hole. The remaining
core of helium and heavy elements quickly accretes into the hole.  The
baryonic mass of the collapsed remnant is 35.9 \Msun \ for Model T90A
and 37.1 \Msun \ for Model T90B, though a few tenths of a solar mass
should be subtracted because of the neutrino losses during the
protoneutron star phase \citep{Oco11}.

The velocity structure at the time when the iron core of T90A
collapses is shown in \Fig{t90coll}. Despite the collapse of the core
to a black hole, the shock waves from the pulses propagate into the
hydrogen envelope long after the core has collapsed, and ultimately
eject it with a kinetic energy $\sim 5 \times 10^{50}$ erg
(\Tab{tmodels}). This leads to a rather ordinary Type IIp supernova
lasting about 200 days with a luminosity on the plateau of about $2
\times 10^{42}$ erg s${-1}$.

Adding rotation to the model does not change the qualitative result,
though it does alter the main sequence mass required to produce the
given black hole mass. Model R70 produces a helium core of 37.4 \Msun
\ that also encounters the pulsational-pair instability and ejects its
hydrogen envelope and a portion of the core. In the end, a remnant
with baryonic mass 34.5 \Msun \ is left and the hydrogen envelope is
ejected producing a Type IIp supernova. Following the transport
of angular momentum that happens during carbon, neon, and oxygen
burning and in the pulses, including magnetic torques, gives an
angular momentum in the collapsing core of less than 10$^{16}$ cm$^2$
s$^{-1}$ in the inner 20 \Msun. Even at the edge of the helium core, j
is only $4 \times 10^{16}$ cm$^2$ s$^{-1}$. The core collapses to a
black hole and does not form a disk.

The evolution of the lighter component, as exemplified by Model T70A,
is similar, but less explosive. This star too encounters a weak
pulsational-pair instability in its oxygen shell during the waning
hour of its life, but the pulses are too weak to eject even the
loosely bound envelope. The 30.1 \Msun \ helium core collapses directly
to a black hole leaving most of its envelope still bound. In a single
star system without rotation, the envelope too would accrete over a
period of days, finally producing a black hole of 55.5 \Msun
\ (\Tab{tmodels}).

\subsection{Binary Evolution}

Now consider these two stars interacting in a binary system with
initial separation of order several AU. On the main sequence, the
stellar radii are sufficiently small, $8.4 \times 10^{11}$ cm and $9.8
\times 10^{11}$ cm for Models T70A and T90A, that the stars evolve
individually. When T90A burns helium however, it fills its Roche lobe
and starts to spill over onto Model T70A. In solitude, T90A
develops a radius of 0.5, 1.0, 3.0, and 6.0 AU when its central helium
mass fraction is 0.978, 0.923, 0.5, and 0.01 respectively. It finally
reaches a radius of 11 AU after helium depletion when the star has
only a few thousand years to live.

The subsequent evolution, especially through one or more stages of
common envelope, is very uncertain \citep{Iva13}, but might resemble
the system described by \citet{Bel16b}. Their initial binary contains
two stars of 96.2 and 60.2 \Msun \ with metallicity 3\% \Zsun \ (their
progenitors were born in the early universe) separated by 11.4 AU. A
stage of Roche lobe overflow strips the primary of its envelope
leaving a helium core of 39 \Msun, similar to models T90AB and R70
here. Half of the envelope is lost from the system. That helium core
collapses to a 35.1 \Msun \ black hole following a loss of 10\% of the
mass to neutrino emission. Later evolution of the secondary leads to a
common envelope that brings the black hole and the helium core of the
secondary into close proximity after the envelope is ejected. The
core of the secondary collapses to a second black hole of 30.8 \Msun
\ in a nearly circular orbit with separation 0.22 AU. The black holes
merge 10.3 Gy later.

Several differences exist with the models in this paper. First, the
models here have higher metallicity and may lose appreciable mass loss
besides mass exchange. Given the key role of the helium cores, the
envelope masses may not be so critical, but they do affect the
parameters of the later common envelope evolution. The slow expansion
of Model T90 to supergiant proportions, likely a consequence of the
way semiconvection is treated in the code, may require a closer
initial orbital separation than assumed by \citet{Bel16b} to transfer
its entire envelope to the secondary. This too may not matter much
because Model T90A becomes a supernova before making a black hole and
ejects any remaining envelope. The orbital separation when the
first black hole forms would be affected though.

Both black holes are formed here by the collapse of iron cores near
2.6 \Msun. There will of necessity be a brief phase of proto-neutron
star formation and neutrino emission before a black hole of about 3
\Msun \ forms. The neutrino losses from that stage are likely to be
only a few tenths of a solar mass \citep{Oco11}, and the efficiency for
neutrino emission declines greatly after the event horizon forms
\citep{Woo86}.  The mass decrement may thus not be as large as the 10\%
assumed by \citet{Bel16b}. Black hole kicks may also be smaller.

A common envelope phase is necessary to bring the black hole and core
of the secondary close enough to merge in a reasonable time. The
expansion history of T70A is thus important. Assuming its core
structure is not greatly altered by the mass accreted during the first
Roche lobe overflow, T70A ignites helium burning 55,000 years after
birth of the first black hole (death of T90A) and expands to 1 AU
110,000 years later, when the helium is half burned. At helium
depletion, 210,000 years after that, the radius is 6.5 AU. Sometime
in between, a common envelope presumably forms, but how close will that
envelope bring the black hole to the helium core before it too
collapses? Estimates typically employ a comparison of envelope binding
energies and core separations plus some efficiency factor $\alpha$
\citep{Iva13}.  A final separation of less than 0.2 AU, like
\citet{Bel16b} require and obtain, implies a gravitational potential
for two $\sim$30 \Msun \ masses of about $8 \times 10^{49}$ erg. The 
{\sl net} binding energy outside 0.2 AU in the T70A pre-supernova
star is only $3 \times 10^{48}$ erg, but additional energy is
expended ejecting matter pushed up from beneath during the common
envelope phase and providing it with ejection speed. The total net
binding energy of the matter outside the hydrogen helium discontinuity
in Model T70A at 31.0 \Msun is indeed large, $2.3 \times 10^{50}$ erg.

\section{Conclusions}

The characteristics of GW 150914 are unlikely to be explained by any
single star model. The model before the event \citep{Fry01} errs in
neglecting the magnetic coupling between the rapidly rotating helium
core and the nearly stationary hydrogen envelope and in neglecting
mass loss. The model after the event \citep{Loe16} errs in invoking an
extremely rapidly rotating helium core for which there is no clear
path in current stellar evolution. An alternative model based on
chemically homogeneous evolution shows promise until mass loss is
included. This model might work if there were no mass loss and could
potentially occur in extremely low metallicity stars - $\sim$10$^{-3}$
\Zsun, but not in stars expected in the galaxy where the merger was
detected.

Similar considerations apply to models based on merger of massive
stars prior the black hole formation. Though, in principle, a rapidly
rotating helium core could be formed, so long as an envelope remained,
magnetic torques would rapidly brake its rotation. If the envelope
were ejected, mass loss during helium burning would brake the core.

In any case, one of the key motivations for single star models, namely
explaining the putative GRB temporally coincident with GW 150914, is
not satisfied.  The Fermi/GBM team reported a possible transient about
0.4 s after the reported LIGO burst trigger time, and lasting for
about one second \citep{Bla15,Con16}, though that detection is head of
controversial.  Studies by \citet{Zha04} have shown that the the GRB
producing jet travels significantly slower than the speed of light
while inside the star that makes it. The jet has far less mass than
the star it working surface, where the jet pushes aside stellar
matter, moves at only about c/3 and thus takes about 10 s to exit a
star with radius $8 \times 10^{10}$ cm in the Zhang study. The most
optimistic model here, R150A has a radius of $6 \times 10^{10}$ cm
when it dies and a much greater density than the model studied by
Zhang et al. The observed GRB here was a weak short one. Thus it is
quite unlikely that the jet would reach the surface in less than 10 s.
The GW signal, on the other hand, reaches the surface in 2 s and
remains forever 8 s ahead of the GRB. That this exceeds the observed
delay by a factor of 10 is a severe problem that should be kept in
mind for future studies of coincident GRBs and GW signals. A repeat of
Zhang's calculation using, e.g., model R150A, would be useful to
clarify the expected delay.  If the star still had an envelope, there
would have been no GRB. Alternatively, it might be possible to make a
prompt GRB by merging two black holes if one or both have a fossil disk
\citep{Per16}.

Indeed, the most likely model for GW 150914 is the one suggested
beforehand. Two massive stars make two black holes that, under the
influence of first, a common envelope, and later, gravitational
radiation, come together long after their creation. The evolution of
rotating and non-rotating stars that might produce black holes with
the measured masses were considered here and found to hover on the
edge of the pair instability. The more massive primary especially is
capable of ejecting its hydrogen envelope owing to thermonuclear
pulses without binary mass transfer, though that does not preclude the
transfer happening. Both stars produce iron cores in hydrostatic
equilibrium with masses of about 2.6 \Msun \ that collapse first to
proto-neutron stars, and emit a few tenths of a solar mass in neutrinos
before becoming black holes into which the rest of the core
accretes. Given the small mass lost, large kicks seem unlikely.

In the future it is hoped that realistic simulations of presupernova
evolution in this mass range can be coupled to models for common
envelope evolution to better describe the progenitor of this
fascinating event.

\section{Acknowledgements}

This research has been partially supported by NASA (NNX14AH34G). The
author appreciates the long term assistance of Alex Heger with the
development of the  \texttt{KEPLER} code and helpful conversations with 
Chris Belczynski about the nature of common envelope evolution.

\end{document}